\documentclass[a4paper]{jpconf}
\usepackage{graphicx}
\usepackage{mathtools}
\usepackage{subfloat}
\usepackage{subfig}
\usepackage{amsmath}
\usepackage{xcolor}
\usepackage{dsfont}
\usepackage{booktabs}
\usepackage{bbold}
\usepackage[utf8]{inputenc}
\usepackage[ngerman, english]{babel}
\usepackage{hyperref}
\numberwithin{equation}{section}

\begin{document}


\title{Lattice investigation of an inhomogeneous phase of the $2+1$-dimensional Gross-Neveu model in the limit of infinitely many flavors}

\author{Marc Winstel, Jonas Stoll and Marc Wagner }{\tiny }

\address{Institut für Theoretische Physik, Goethe Universität Frankfurt am Main, 60438 Frankfurt am Main, Germany}

\ead{winstel@itp.uni-frankfurt.de, stoll@itp.uni-frankfurt.de, mwagner@th.physik.uni-frankfurt.de }

\begin{abstract}
We investigate the phase structure of the $2+1$ dimensional Gross-Neveu model in the large-$N_f$ limit, where $N_f$ denotes the number of fermion flavors. We discuss two different fermion representations and their implication on the interpretation of a discrete symmetry of the action. We present numerical results, which indicate the existence of an inhomogeneous phase similar as in the 1+1-dimensional Gross-Neveu model.     
\end{abstract}

\section{Introduction, $2+1$-dimensional Gross-Neveu model in the large-$N_f$ limit}

The calculation of the full phase diagram of QCD, either analytically or numerically, is a very challenging and still unsolved problem. In particular lattice QCD is at the moment restricted to small chemical potential, because of the sign problem (see e.g \cite{Philipsen:2010gj} and references therein). Therefore, it is of interest to study the phase structure of simpler quantum field theories, which share some features with QCD and, thus, might serve as crude models or even effective theories for QCD. A common and very simple example is the Gross-Neveu (GN) model in $1+1$ spacetime dimensions \cite{Gross:1974jv}, which describes $N_f$ fermion flavors and exhibits a discrete chiral symmetry, which can be spontaneously broken. In the large-$N_f$ limit an inhomogeneous phase was found, i.e.\ a phase where the chiral order parameter is a periodic function of the spatial coordinate \cite{Thies:2003kk,Schnetz:2004vr}. Recently, numerical evidence was presented, that a similar inhomogeneous phase also exists at finite $N_f$ \cite{Pannullo:2019bfn}.
 
As an intermediate step towards $3+1$ spacetime dimensions we consider in this work the GN model in $2+1$ dimensions in the large-$N_f$ limit. The ``homogeneous phase diagram'', where the chiral condensate is assumed to be constant, was calculated in ref.\ \cite{Rosenstein:1988dj}. In ref.\ \cite{Urlichs:2007zz} a particular stripe-like 1-dimensional modulation of the condensate has been investigated. It turned out that it is not favored over a spatially constant condensate. Here we explore the phase structure numerically using lattice field theory with particular focus on the possible existence of an inhomogeneous phase. The Euclidean action and the partition function are
\begin{align}
S[\bar{\psi},\psi] = \int d^3x \, \bigg(\sum_{n=1}^{N_f} \bar\psi_n \Big(\gamma_\nu \partial_\nu + \gamma_0 \mu\Big) \psi_n - \cfrac{g^2}{2} \bigg(\sum_{n=1}^{N_f} \bar\psi_n \psi_n\bigg)^2\bigg) \quad , \quad Z = \int D\bar{\psi} \, D\psi \, e^{-S[\bar{\psi},\psi]}. 
\end{align}
$\psi_n$ denotes a fermionic field with flavor index $n$, $\mu$ is the chemical potential and $g$ is the coupling constant. Representations for the spin matrices $\gamma_0$, $\gamma_1$ and $\gamma_2$ are discussed in section~\ref{SEC002}. Introducing a real scalar field $\sigma$ and performing the integration over the Grassmann-valued fermionic fields one obtains the equivalent effective action and partition function
\begin{align}
\label{S_eff} S_{\text{eff}}[\sigma] = N_f \bigg(\frac{1}{2 \lambda} \int d^3x \, \sigma^2 - \ln\Big(\det(Q)\Big)\bigg) \quad , \quad Z = \int D\sigma \, e^{-S_{\text{eff}}[\sigma]} ,
\end{align}
where $\lambda = N_f g^2$ and $Q = \gamma_\nu \partial_\nu + \gamma_0 \mu + \sigma \mathbb{1}$ ($\nu = 0,1,2$ is a spacetime index, $\mathbb{1}$ denotes an identity matrix in spinor space). In this work we restrict the dependence of $\sigma$ to the spatial coordinates, i.e.\ either $\sigma = \sigma(x)$ (see section~\ref{SEC_X}) or $\sigma = \sigma(x,y)$ (see section~\ref{SEC_XY}). Then one can show that $S_{\text{eff}}[\sigma]$ is real, which is essential for numerical calculations. Moreover, because of the factor $N_f \rightarrow \infty$ on the right hand side of $S_{\text{eff}}[\sigma]$, only field configurations $\sigma$ corresponding to a global minimum of $S_{\text{eff}}[\sigma]$ contribute to the partition function $Z$.


\section{\label{SEC002}Discrete symmetry $\sigma \rightarrow -\sigma$ and fermion representations}

One can show that the effective action in eq.\ (\ref{S_eff}) has a discrete symmetry $\sigma \rightarrow -\sigma$, i.e.\ $S_{\text{eff}}[\sigma] = S_{\text{eff}}[-\sigma]$. Moreover, $\sigma \propto \langle \sum_{n=1}^{N_f} \bar\psi_n \psi_n \rangle$.

In $1+1$ spacetime dimensions a possible irreducible $2 \times 2$ representation for the $\gamma$ matrices is
\begin{align}
\gamma_0 = \sigma_1 \quad , \quad \gamma_1 = \sigma_2 ,
\end{align}
where $\sigma_j$ denote the Pauli matrices. A non-vanishing $\sigma$ would indicate spontaneous breaking of the symmetry
\begin{align}
\label{EQN569} \psi_n \rightarrow \sigma_3 \psi_n .
\end{align}
Since $\sigma_3$ anticommutes with $\gamma_0$ and $\gamma_1$, it is appropriate to define $\gamma_5 = \sigma_3$ and to interpret the symmetry (\ref{EQN569}) as discrete chiral symmetry.

A possible irreducible $2 \times 2$ representation for the $\gamma$ matrices in $2+1$ spacetime dimensions is
\begin{align}
\label{EQN432} \gamma_0 = \sigma_1 \quad , \quad \gamma_1 = \sigma_2 \quad , \quad \gamma_2 = \sigma_3 .
\end{align}
One can show that it is impossible to find a corresponding appropriate $\gamma_5$ matrix, i.e.\ a matrix, which anticommutes with $\gamma_0$, $\gamma_1$ and $\gamma_2$. Consequently, a non-vanishing $\sigma$ cannot be interpreted as an indication for chiral symmetry breaking. A possibility to retain the interpretation of $\sigma$ as chiral order parameter is to use a reducible $4 \times 4$ representation for the $\gamma$ matrices,
\begin{align}
\label{EQN852} \gamma_0 = \left(\begin{array}{cc}
+\sigma_1 & 0 \\
0 & -\sigma_1
\end{array}\right) \quad , \quad
\gamma_1 = \left(\begin{array}{cc}
+\sigma_2 & 0 \\
0 & -\sigma_2
\end{array}\right) \quad , \quad
\gamma_2 = \left(\begin{array}{cc}
+\sigma_3 & 0 \\
0 & -\sigma_3
\end{array}\right) .
\end{align}
For a detailed discussion of fermion representations in $2+1$ dimensions and its implicatons we refer to refs.\ \cite{Appelquist:1986fd,Scherer:2012nn}.


\section{Numerical results}

In this section we explore the phase structure of the $2+1$-dimensional GN model in the large-$N_f$ limit numerically. We have performed the majority of computations using the $2 \times 2$ representation (\ref{EQN432}) for the $\gamma$ matrices, but also some computations using the $4 \times 4$ representation (\ref{EQN852}). Within numerical precision the results are in agreement. All results shown in the following were obtained using the $2 \times 2$ representation (\ref{EQN432}). Based on known analytical results for the phase diagram of the $1+1$-dimensional GN model \cite{Thies:2003kk,Schnetz:2004vr} we expect the following three phases:
\begin{itemize}
\item a symmetric phase, characterized by $\sigma = 0$;

\item a homogeneously broken phase, characterized by $\sigma = \textrm{const} \neq 0$;

\item an inhomogeneous phase, where $\sigma = \sigma(x,y)$ depends on the spatial coordinates.
\end{itemize} 
The spactime volume in our computations is finite with temporal extent $1/T$, where $T$ is the temperature, and spatial volume $L^2$. We discretize the effective action (\ref{S_eff}) using a plane wave expansion for the temporal direction and lattice field theory for the two spatial directions, where we decided for naive fermions for the fermionic determinant (for a discussion of naive fermions see standard textbooks on lattice field theory, e.g. ref. \cite{Rothe:1992nt}). Technical aspects are similar as in refs.\ \cite{deForcrand:2006zz,Wagner:2007he,Heinz:2015lua} and will be discussed in detail in an upcoming publication.


\subsection{$\sigma = \textrm{const}$}

\begin{figure}[htb]
\begin{center}
\includegraphics[width=7.0cm]{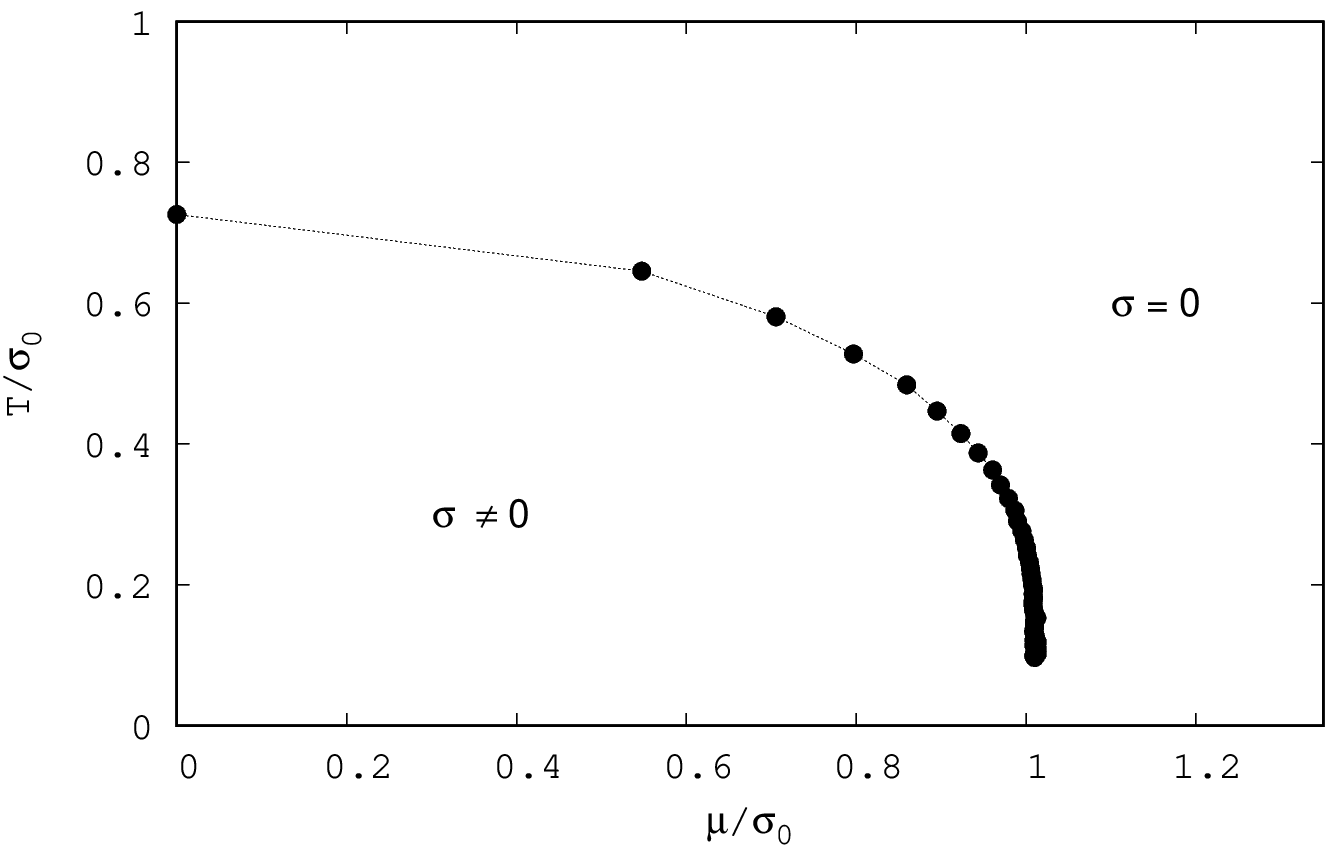}	
\end{center}
\caption{\label{hompd}Phase diagram of the $2+1$-dimensional GN model in the $\mu$-$T$ plane for $\sigma = \textrm{const}$ (black dots are our numerical results, while the dashed line is just a guide to the eye).}
\end{figure}

In a first step we determined the phase diagram for homogeneous $\sigma$, i.e.\ not allowing any spatial modulation for $\sigma$. As a byproduct we obtained $\sigma_0 = \sigma|_{\mu=0,T=0}$, which we use to set the scale. It is known that there is a symmetric phase and a homogeneously broken phase \cite{Rosenstein:1988dj}. We determined the boundary between these two phases by numerically minimizing $S_{\text{eff}}$ with respect to the constant $\sigma$ using a standard algorithm for 1-dimensional minimization (a variant of the golden section search). Our result, which is shown in Fig.\ \ref{hompd}, is consistent with the result from ref.\ \cite{Rosenstein:1988dj}.

\subsection{\label{SEC_X}$\sigma = \sigma(x)$}

In a second step, we allowed $\sigma$ to depend on one of the two spatial coordinates, i.e.\ $\sigma = \sigma(x)$. Then $S_{\text{eff}}$ depends on $N_s$ variables $\sigma_j$, which represent $\sigma(x)$ on the lattice sites $x = j a$, where $a = L / N_s$ denotes the lattice spacing. Since finding the global minimum of a function in many variables is a very challenging task, we started by performing stability analyses with respect to $\sigma = 0$. This amounts to finding the eigenvalues and eigenvectors of the Hessian matrix
\begin{equation}
\label{EQN777} H_{j k} = \frac{\partial^2}{\partial \sigma_j \partial \sigma_k} S_{\text{eff}}\Big|_{\sigma_0 = \sigma_1 = \ldots = \sigma_{N_s-1} = 0} ,	 
\end{equation}
where negative eigenvalues indicate directions, in which $S_{\text{eff}}$ decreases.

\begin{figure}[htb]
\begin{center}
\includegraphics[width=7.cm]{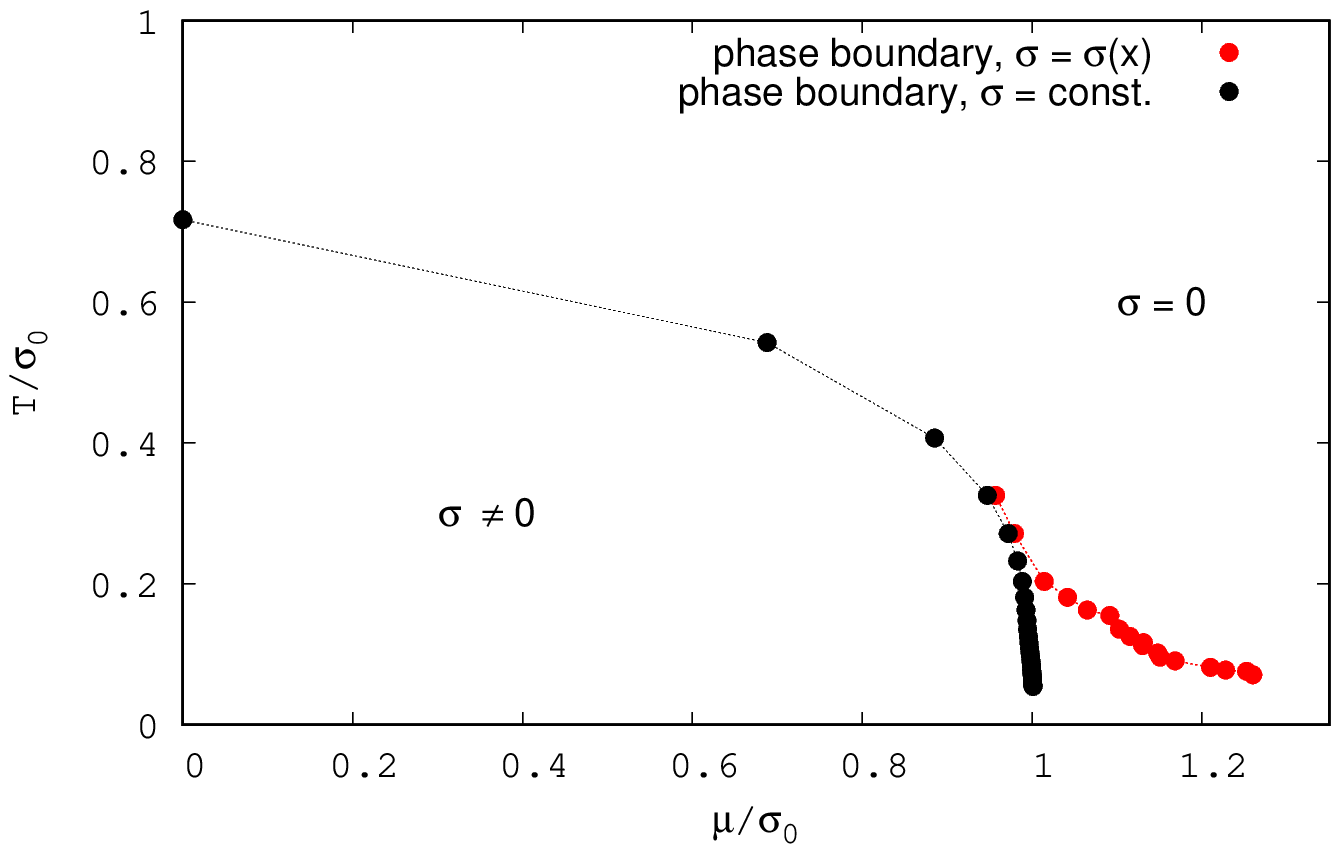}
\hfill
\includegraphics[width=7.cm]{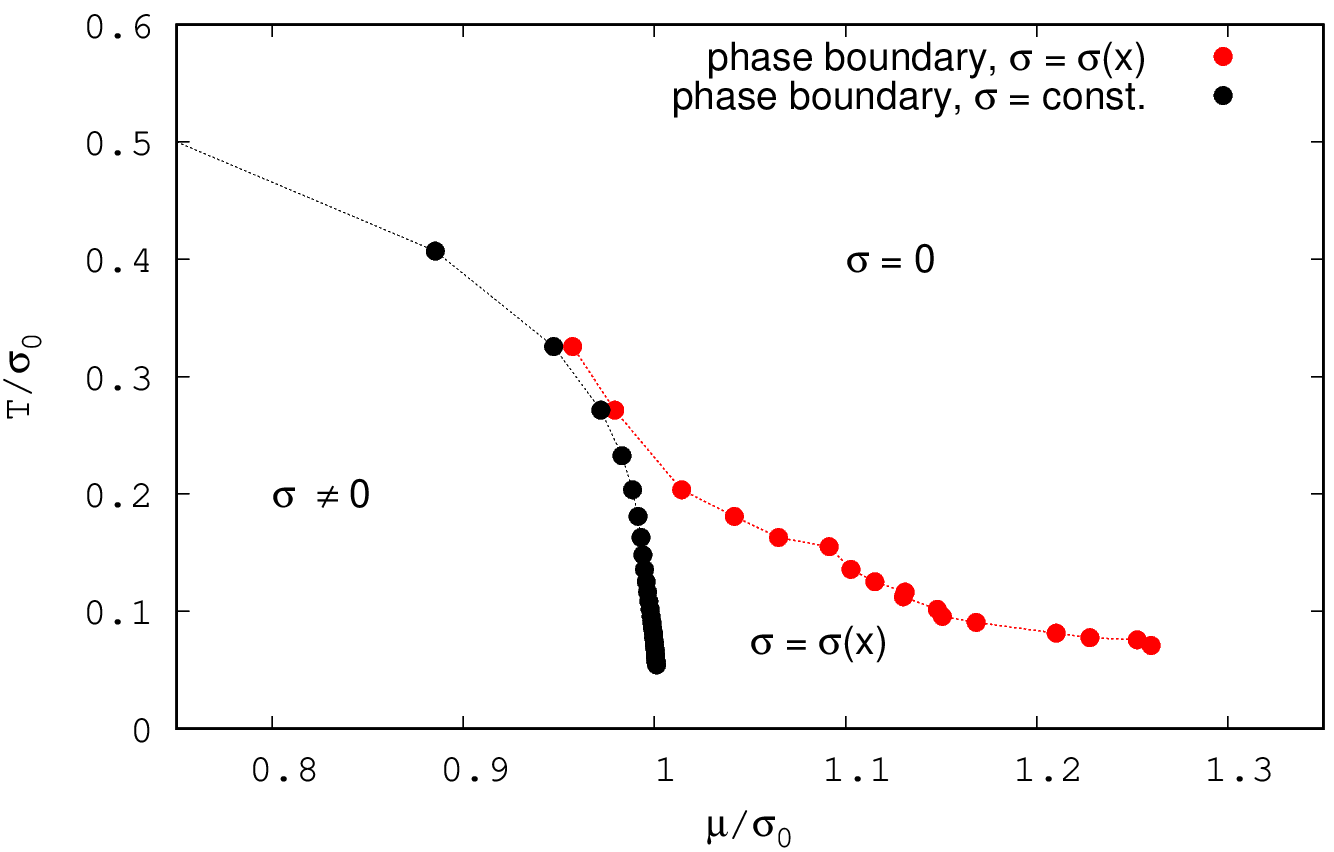}
\end{center}
\caption{\label{onedimpd}``Phase diagram'' of the $2+1$-dimensional GN model in the $\mu$-$T$ plane for $\sigma = \sigma(x)$. The black dots represent the phase boundary for $\sigma = \textrm{const.}$ also shown in Fig.\ \ref{hompd}. The red dots are obtained by stability analyses with respect to $\sigma = 0$. The triangular region between the red dots and the black dots is part of an inhomogeneous phase. The right plot is a zoomed version of the left plot.}
\end{figure}

\begin{figure}[htb]
\begin{center}
\includegraphics[width=5.5cm]{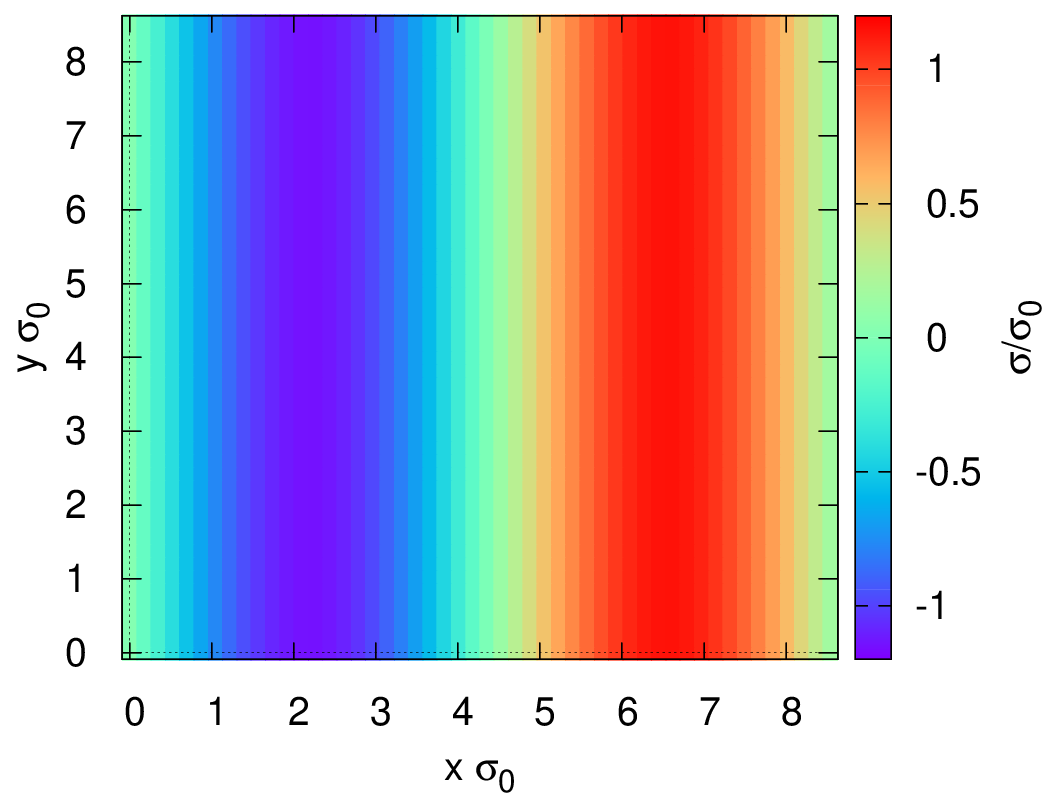} 
\hfill
\includegraphics[width=5.5cm]{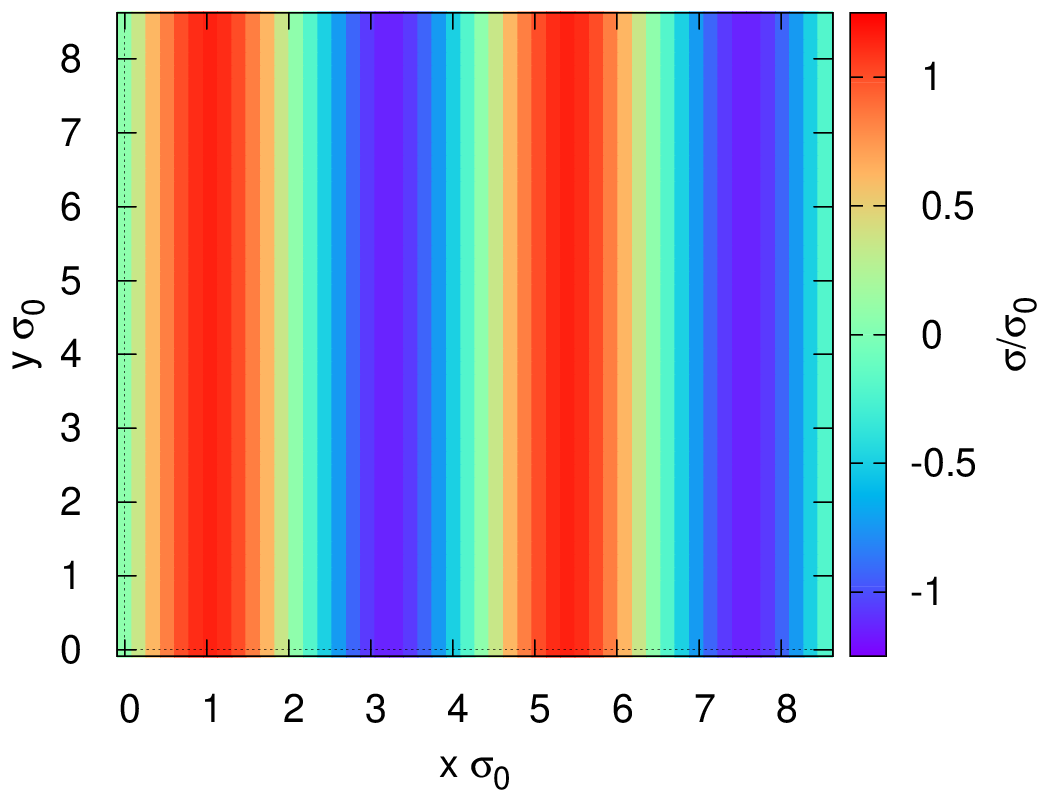} 
\end{center}
\caption{\label{evs}Eigenvectors of the Hessian matrix (\ref{EQN777}) corresponding to negative eigenvalues. \textbf{(left)}~$\mu/\sigma_0 = 1.011$, $T/\sigma_0 = 0.198$. \textbf{(right)}~$\mu/\sigma_0 = 1.064$, $T/\sigma_0 = 0.118$.}
\end{figure}

The red dots in Fig.\ \ref{onedimpd} separate a region, where $\sigma = 0$ is stable (i.e.\ no negative eigenvalues of $H$), from another region, where $\sigma = 0$ is not stable (i.e.\ at least one negative eigenvalue of $H$). Note that the red dots do not necessarily correspond to a phase boundary (even though they could coincide with a phase boundary, as it is the case e.g.\ for the $1+1$ dimensional GN model, where they separate the symmetric phase and the inhomogeneous phase). Nevertheless, they unambiguously signal the existence of an inhomogeneous phase, which covers the triangular region between the red dots and the black dots (the latter represent the phase boundary for $\sigma = \textrm{const}$ already shown in Fig.\ \ref{hompd}). This inhomogeneous phase might, however, have a larger extension, which we are currently investigating by performing full multi-dimensional minimizations of $S_{\text{eff}}$ using a conjugate gradient algorithm. Note that the red dots do not seem to correspond to a smooth curve. This could be a finite-volume effect, similar to that observed in a lattice field theory study of the $1+1$ dimensional GN model \cite{deForcrand:2006zz}.
In Fig.\ \ref{evs} we show two examples of eigenvectors of the Hessian matrix (\ref{EQN777}) corresponding to negative eigenvalues. These eigenvectors indicate the shape of spatial modulations of $\sigma(x)$, which lower the effective action compared to $\sigma = 0$, i.e.\ $S_\textrm{eff}[\sigma(x)] < S_\textrm{eff}[0]$. The wave number increases for increasing chemical potential and fixed temperature, a behavior also observed for the $1+1$-dimensional GN model \cite{Thies:2003kk,Schnetz:2004vr}.


\subsection{\label{SEC_XY}$\sigma = \sigma(x,y)$}

\begin{figure}[htbp]
\begin{center}
\includegraphics[width=5.5cm]{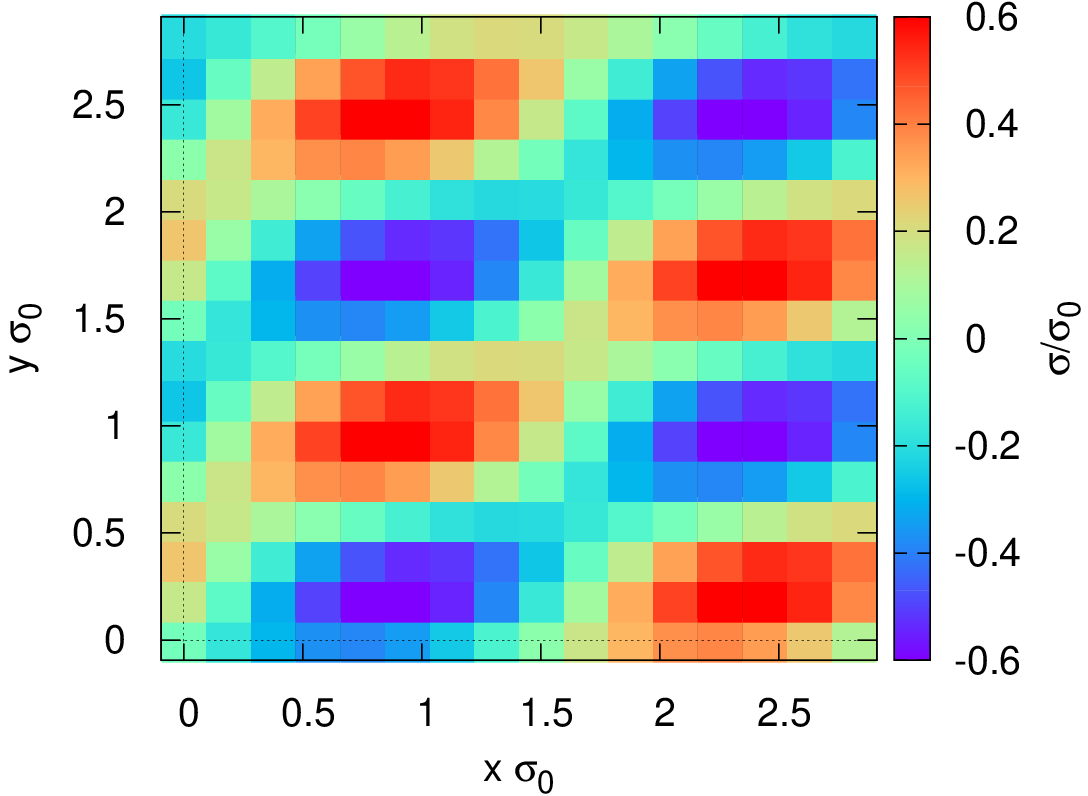}
\end{center}
\caption{\label{evs2d}Eigenvector of the Hessian matrix corresponding to a negative eigenvalue. $\mu/\sigma_0 = 5.5$, $T/\sigma_0 = 0.19$.}
\end{figure}

We also started to explore the phase structure for $\sigma$ depending on both spatial coordinates, i.e.\ $\sigma = \sigma(x,y)$. We are not yet in a position to show a phase diagram. It is, however, worthwhile to note that we found negative eigenvalues of the Hessian matrix corresponding to eigenvectors, which oscillate both in $x$ and in $y$ direction (see Fig.\ \ref{evs2d} for an example). This might indicate that the inhomogeneous phase is larger than in Fig.\ \ref{onedimpd}. Note, however, that these first results have been obtained on a rather small lattice with very coarse lattice spacing and might, thus, suffer from sizeable finite volume corrections and discretization errors. We plan to perform similar, but more precise computations on larger and finer lattices in the near future. It will also be interesting to explore, whether there are regions in the $\mu$-$T$ plane, where the global minimum of $S_{\text{eff}}$ is related to such 2-dimensional structures of $\sigma(x,y)$.

\section*{Acknowledgments}

We thank J.\ Braun, A.\ Königstein and A.\ Wipf for important and helpful discussions on chiral symmetry and its relation to fermion representations in $2+1$ spacetime dimensions.
M.\ Winstel thanks the organizers of the ``FAIRness 2019'' conference for the opportunity to give this talk. 
J.\ Stoll, M.\ Winstel and M.\ Wagner acknowledge support by the Deutsche Forschungsgemeinschaft (DFG, German Research Foundation) through the CRC-TR 211 ``Strong-interaction matter under extreme conditions'' -- project number 315477589 -- TRR 211.
M.\ Wagner acknowledges support by the Heisenberg Programme of the Deutsche Forschungsgemeinschaft, grant WA 3000/3-1.
This work was supported in part by the Helmholtz International Center for FAIR within the framework of the LOEWE program launched by the State of Hesse. 
Calculations on the GOETHE-HLR and on the FUCHS-CSC high-performance computer of the Frankfurt University were conducted for this research. We would like to thank HPC-Hessen, funded by the State Ministry of Higher Education, Research and the Arts, for programming advice.



\end{document}